\documentclass[preprintnumbers,twocolumn,amsmath,amssymb,amsfonts,superscriptaddress,floatfix,showpacs,prd,aps]{revtex4}
\newcommand{\del}{\partial}
\newcommand{\Tr}{\mathrm{Tr}}

\newcommand{\vev}[1]{\langle{#1}\rangle}

\usepackage{graphicx,color,mathrsfs,flafter
}

\begin{document}
\title{ Phase diagram in an external magnetic field beyond a mean-field approximation}

\author{V. Skokov} \email[E-Mail:]{VSkokov@bnl.gov} \affiliation{%
Physics Department, Brookhaven National Laboratory,
Upton, NY 11973, USA
	}



\begin{abstract}
  The phase structure of the Polyakov loop-extended
  chiral quark--meson  model is explored in a nonperturbative approach, 
	beyond   a   mean-field approximation, in the presence of a magnetic field. 
	We show that by including meson fluctuations one cannot resolve the qualitative 
	discrepancy on the dependence of the crossover transition temperature in a non-zero magnetic field
	between effective model predictions 
	and recent lattice results (arXiv:1111.4956).
	 We compute the curvature of the crossover line in the $T-\mu_B$ plane at
	 a non-zero magnetic field and show that the curvature increases with increasing magnetic field.    
	On the basis of QCD inequalities,  we  also argue that, at least in the 
	large $N_c$ limit, a chiral critical end point and, consequently, 
	a change from crossover to a first-order chiral phase transition are excluded at zero baryon chemical potential 
	and non-zero magnetic field. 
\end{abstract}

\preprint{BNL-96687-2011-JA}

\maketitle

\section{Introduction}

The phase diagram of strongly interacting matter in the presence of an 
external magnetic field has been been explored
in model studies and in first principle  Lattice Quantum Chromodynamics (LQCD) 
calculations.  
These investigations has an importance for physics of heavy-ion collisions 
at top RHIC and LHC energies, where according to the estimates 
(see Refs.~\cite{Kharzeev:2007jp} and~\cite{Skokov:2009qp}),  the magnitude of the
magnetic field may reach extremely large values of  $eB \sim 10  m_\pi^2$.   

Model studies (see e.g.
Refs.~\cite{Johnson:2008vna,Fraga:2008um,Mizher:2010zb,arXiv:1012.1291,arXiv:1007.0790})
have revealed a general structure of the phase diagram in the temperature-magnetic field ($T-eB$) plane. 
They showed that  
the chiral crossover temperature $T_{\rm pc}$ increases with increasing magnetic field. 
This result has been, however,  obtained within a mean-field approximation, i.e.
meson fluctuations have not been taken into account. However, charged pion
degrees of freedom interacting with the magnetic field 
may  drastically change properties of the crossover transition even at moderate values of $B$. 

Early LQCD calculations~\cite{D'Elia:2010nq} confirmed  the increase of $T_{\rm pc}$ with $eB$.   
Unfortunately,  results of Ref.~\cite{D'Elia:2010nq} may not be final: 
the calculations were performed with standard staggered fermions, large lattice spacing 
and with somewhat heavy pion masses. These deficiencies may  diminish the role of charged pions.  

Recent LQCD studies~\cite{Bali:2011qj} at physical pion mass, with 
improved staggered fermions and extrapolation to the continuum
have shown completely opposite  dependence of $T_{\rm pc}$ on the magnetic field $eB$. It has been found that the 
transition temperature  significantly decreases with increasing magnetic field. 

In this article,  by analysing the chiral structure of the phase diagram 
of the Polyakov loop-extended quark-meson model 
beyond a mean-field approximation, we  demonstrate, that  
meson fluctuations cannot explain above mentioned qualitative discrepancy in the behavior of the phase
transition temperature with the magnetic field.  

Some model calculations have also  predicted that at large values 
of the magnetic field the strength of the transition  
increases and eventually, at a critical end point, 
the chiral crossover turns into a first-order phase transition~\cite{Fraga:2008um}. 
Our model calculations with physical number of colors  $N_c=3$ 
show no evidence in favour of possible chiral 
first-order phase transition in a wide range of the magnetic field. 
Based on the no-go theorem formulated for QCD in Ref.~\cite{arXiv:1110.3044}, 
we will also argue that a chiral critical end point and  a change from chiral crossover 
to a chiral first-order phase transition 
are forbidden at the leading order of the 
large $N_c$ expansion for any non-zero pion mass.  

In this paper, we do not consider the, so-called, no-sea approximation, the essence of which
boils down to neglecting   vacuum contributions in a thermodynamic potential.
As was shown analytically  in Ref.~\cite{MFonVT}, the no-sea approximation 
results in a first-order chiral phase transition at zero chemical potential
in the chiral limit (already on the mean-field level). 
This is in direct contradiction with  recent LQCD findings~\cite{arXiv:0909.5122}. 
We refer to Refs.~\cite{Mizher:2010zb} and~\cite{MFonVT}
for the reader interested in a comparison between results of a properly renormalized theory 
and those obtained in the no-sea approximation at finite and zero magnetic field correspondingly.  

The structure of the paper is as follows. In the next section we discuss  the 
functional renormalization  group  flow equations in the  non-trivial magnetic background.
In Section 3, we review the mean-field approximations for the Polyakov loop-extended Quark-Meson (PQM)  model. 
Since the PQM model is renormalizable, we perform dimensional regularization 
to subtract divergences arising from the vacuum term. 
Section 4 is devoted to the main results and discussions, including above mentioned 
input from the large $N_c$ limit. Section 5 contains our conclusions.

\section{The Polyakov loop-extended quark-meson model}\label{sec:pqm}

The quark--meson model is an effective realization of the low--energy
sector of QCD. The model is built to respect symmetries of QCD, including the chiral symmetry
in the limit of vanishing pion mass. 
However, because the local color $SU(N_c)$ invariance of QCD is replaced by a global
symmetry, the model does not describe confinement.
The improved version of the model, the so-called Polyakov loop-extended quark-meson (PQM) model 
incorporates a coupling of the quarks to a uniform temporal
color gauge field, represented by the Polyakov loop. In the PQM model,  many facets of confinement
can be emulated~\cite{Fukushima, PNJL,Schaefer:PQM}.

The Lagrangian of the PQM model reads
\begin{eqnarray}\label{eq:pqm_lagrangian}
  {\cal L} &=& \bar{q} \, \left[i\gamma^\mu {D}_\mu  - g (\sigma + i \gamma_5
    \vec \tau \vec \pi )\right]\,q \  -{\cal U}({\Phi},{\Phi}^{*}) \\&&+ 
  \frac 1 2 (\partial_\mu \sigma)^2 + \frac{ 1}{2}
  (\partial_\mu  \pi^0)^2 + {\cal D}_\mu \pi^+ {\cal D}^{\mu} \pi^- 
    - U(\sigma, \vec \pi ). \nonumber
\end{eqnarray}
The coupling between the effective gluon field and quarks, and between the (electro)magnetic field $B$
and quarks  is implemented through the covariant derivative
\begin{equation}
  D_{\mu}=\del_{\mu}-iA_{\mu} - i Q A^{\rm EM}_\mu,
\end{equation}
where $A_\mu=g\,A_\mu^a\,\lambda^a/2$ and $A^{\rm EM}_\mu=(0,Bx,0,0)$. 
The flavor matrix $Q$ is defined by the quark electric charges $Q={\rm diag}(2/3 e, -1/3 e)$. 
The spatial components of the
gluon field are neglected, i.e. $A_{\mu}=\delta_{\mu0}A_0$. 
The interaction of  charged pion
$\pi^{\pm}=(\pi_1\pm i \pi_2)/\sqrt{2}$ with the electromagnetic field  
is included  by 
${\cal D}_\mu = \partial_\mu + i e A_\mu^{\rm EM}$.  
The effective potential for the gluon
field  ${\cal U}({\Phi},{\Phi}^{*})$ is expressed in terms of the thermal expectation values of the
color trace of the Polyakov loop and its conjugate
\begin{equation}
  {\Phi}=\frac{1}{N_c}\vev{\Tr_c L(\vec{x})},\quad {\Phi}^{*}=\frac{1}{N_c}\vev{\Tr_c
    L^{\dagger}(\vec{x})},
\end{equation}
with
\begin{eqnarray}
  L(\vec x)={\mathcal P} \exp \left[ i \int_0^\beta d\tau A_4(\vec x , \tau)
  \right]\,,
\end{eqnarray}
where ${\mathcal P}$ stands for the path ordering, $\beta=1/T$ and
$A_4=i\,A_0$.  In the $O(4)$ representation, the meson field is
introduced as $\phi=(\sigma,\vec{\pi}=({\pi_0,\pi_1,\pi_2}))$ and  the corresponding
$SU(2)_L\otimes SU(2)_R$ chiral representation is defined by
$\sigma+i\vec{\tau}\cdot\vec{\pi}\gamma_5$.

The meson potential of the model, $U(\sigma,\vec{\pi})$, is
defined as
\begin{equation}
  U(\sigma,\vec{\pi})=\frac{\lambda}{4}\left(\sigma^2+\vec{\pi}
    ^2-v^2\right)^2-c\sigma.
\end{equation}
The effective potential of the gluon field is parametrized in
such a way as to preserve the $Z(3)$ invariance:
\begin{equation}
  \frac{{\cal U}({\Phi},{\Phi}^{*})}{T^4}=
  -\frac{b_2(T)}{2}{\Phi}^{*}{\Phi}
  -\frac{b_3}{6}({\Phi}^3 + {\Phi}^{*3})
  +\frac{b_4}{4}({\Phi}^{*}{\Phi})^2\,\label{eff_potential}.
\end{equation}
The parameters,
\begin{eqnarray}
  \hspace{-4ex}
  b_2(T) &=& a_0  + a_1 \left(\frac{T_0}{T}\right) + a_2
  \left(\frac{T_0}{T}\right)^2 + a_3 \left(\frac{T_0}{T}\right)^3\,
\end{eqnarray}
with $a_0 = 6.75$, $a_1 = -1.95$, $a_2 = 2.625$, $a_3 = -7.44$, $b_3 =
0.75$,  $b_4 = 7.5$ and $T_{0}=270$ MeV were chosen to reproduce the equation of state
of the pure SU$_c$(3) lattice gauge theory.  When the 
coupling to the quark degrees of freedom are neglected, the potential~(\ref{eff_potential}) yields a
first-order deconfinement phase transition at
$T_0$.

\subsection{The FRG method in the PQM model}\label{sec:rg}

To take fluctuations into account in the PQM
model, we use  the functional renormalization group (FRG) method.
This method is based on  an infrared regularization of the fluctuations at a sliding
momentum scale $k$, resulting in a scale-dependent effective action $\Gamma_k$, 
the so-called effective average action~\cite{Berges:review}.
In this article,  the Polyakov loop
is treated as a background field  introduced self-consistently on the
mean-field level. Quark and meson fluctuations are accounted for by solving the FRG flow equations.

The FRG flow equation for the PQM model  at 
zero magnetic field was derived in Ref.~\cite{Skokov:2010wb}. 
The derivation for finite magnetic field is lengthy, but straightforward
and can be done along similar lines.
Here we provide only the final expression for the flow equation. 
It can be easily proved that in the limit $B\to0$ the flow equation 
reduces to that of  Ref.~\cite{Skokov:2010wb}.
\begin{widetext}
  Following our previous work~\cite{Skokov:2010wb}, 
  the flow equation for the scale-dependent grand canonical potential density, $\Omega_{k}=T\Gamma_{k}/V$, is formulated 
	for the quark and meson subsystems in the leading order of derivative expansion
  \begin{eqnarray}\label{eq:frg_flow}
 &&   \del_k \Omega_k({\Phi}, {\Phi}^*; T,\mu)=\frac{k^4}{12\pi^2}
    \left\{  \Bigg[ \frac{ 1+2n_B(E_\pi;T)}{E_\pi} \Bigg]
      + \Bigg[  \frac{1+2n_B(E_\sigma;T)  }{E_\sigma} \Bigg]   \right\}
 \\ \nonumber && \quad
 + k \frac{eB}{2\pi^2} \sum_{n=0}^{\infty} \sqrt{k^2-q_\perp^2(n,e,0)} \theta(k^2-q_\perp^2(n,e,0))  \Bigg[ \frac{ 1+2n_B(E_\pi;T)}{E_\pi} \Bigg]
 \\ \nonumber &&\quad   -  k \sum_{f=1,2}   \frac{N_c Q_{ff} B}{2 \pi^2}
 \sum_{s=\frac1 2, -\frac 1 2} \sum_{n=0}^{\infty} \sqrt{k^2-q_\perp^2(n,Q_{ff},s)} \theta(k^2-q_\perp^2(n,Q_{ff},s)) 
\Bigg[ \frac{1-
      N({\Phi},{\Phi}^*;T,\mu)-\bar{N}({\Phi},{\Phi}^*;T,\mu)}{E_q}\Bigg] .
  \end{eqnarray}
  Here $n_B(E_{\pi,\sigma};T)$ is the bosonic distribution function
  \begin{equation*}
    n_B(E_{\pi,\sigma};T)=\frac{1}{\exp({E_{\pi,\sigma}/T})-1},
  \end{equation*}
  with the pion and sigma energies
  \begin{equation*}
    E_\pi = \sqrt{k^2+\overline{\Omega}^{\,\prime}_k}\;~,~ E_\sigma
    =\sqrt{k^2+\overline{\Omega}^{\,\prime}_k+2\rho\,\overline{\Omega}^{\,
        \prime\prime} _k},
  \end{equation*}
where the primes denote derivatives with respect to $\rho = (\sigma^2+\vec{\pi}^2)/2$ of
  $\overline{\Omega}=\Omega+c\sigma$.
  The fermion distribution functions $N({\Phi},{\Phi}^*;T,\mu)$ and
  $\bar{N}({\Phi},{\Phi}^*;T,\mu)$,
  \begin{eqnarray}\label{n1}
    N({\Phi},{\Phi}^*;T,\mu)&=&\frac{1+2{\Phi}^*\exp[\beta(E_q-\mu)]+{\Phi} \exp[2\beta(E_q-\mu)]}{1+3{\Phi} \exp[2\beta(E_q-\mu)]+
      3{\Phi}^*\exp[\beta(E_q-\mu)]+\exp[3\beta(E_q-\mu)]},  \\
    \bar{N}({\Phi},{\Phi}^*;T,\mu)&=&N({\Phi}^*,{\Phi};T,-\mu),
    \label{n2}
  \end{eqnarray}
  are modified because of the  coupling to the gluon field. The quark energy is given by
  \begin{equation}
    \label{dispertion}
    E_q =\sqrt{k^2+2g^2\rho}.
  \end{equation}
The function $q^2_\perp$ is defined as $q^2_\perp(n,q,s) = (2n+1-2s) |q| B$  and has the same structure 
as in mean-field approximation, see Ref.~\cite{Mizher:2010zb}.

If one replaces the energies of particles by their tree-level approximation and integrate the flow equation with respect to 
the scale $k$, then, after integration by parts, the effective thermodynamic potential $\Omega$ reduces to the 
one-loop result for the PQM model in the presence of non-zero magnetic field. 
This also proves  validity of the flow equation (\ref{eq:frg_flow}). 

  \end{widetext}

The minimum of the thermodynamic potential is determined by the
stationarity condition
\begin{equation}
  \left. \frac{d \Omega_k}{ d \sigma} \right|_{\sigma=\sigma_k}=\left. \frac{d
      \overline{\Omega}_k}{ d \sigma} \right|_{\sigma=\sigma_k} - c =0.
  \label{eom_sigma}
\end{equation}
We solve the flow equation~(\ref{eq:frg_flow})  numerically with the
initial cutoff $\Lambda=1.2$ GeV (see additional details in Ref.~\cite{Skokov:2010wb}).  
The initial conditions  for the flow are
fixed to reproduce { the  in-vacuum properties ($T=\mu=0$, $eB=0$) :} the physical pion mass $m_{\pi}=138$
MeV, the pion decay constant $f_{\pi}=93$ MeV, the sigma mass
$m_{\sigma}=600$ MeV, and the constituent quark mass $m_q=300$ MeV at
the scale $k\to 0$.  We treat the symmetry breaking term, $c=m_\pi^2 f_\pi$ as 
an external field. 
We also neglect  the flow of the Yukawa coupling $g$, {which is
not expected to be significant for the present studies~(see e.g. Refs.~\cite{Jungnickel,hep-ph/9705474}). }

The thermodynamic potential~(\ref{omega_final}) does not contain
contributions of thermal modes with momenta larger than the cutoff
$\Lambda$.  To obtain the correct high-temperature behavior
of thermodynamic functions we need to supplement the FRG potential with the
contribution of the high-momentum states. For this, we follow the 
procedure described in Ref.~\cite{Skokov:2010wb}: at high 
$k>\Lambda$ the meson contribution to the flow in the equation is disregarded
 and only  
the flow of massless quarks interacting with the Polyakov loop is considered.

By solving Eq.~(\ref{eq:frg_flow}), one obtains the thermodynamic
potential for the quark and meson subsystems, $\Omega_{k\to0} ({\Phi},
{\Phi}^*;T, \mu)$, as a function of the Polyakov loop variables ${\Phi}$
and ${\Phi}^*$. The full thermodynamic potential $\Omega({\Phi}, {\Phi}^*;T,
\mu)$ in the PQM model, including quark, meson and gluon
degrees of freedom, is obtained by adding the effective gluon potential ${\cal U}({\Phi},
{\Phi}^*)$  to $\Omega_{k\to0} ({\Phi}, {\Phi}^*;T, \mu)$:
\begin{equation}
  \Omega({\Phi}, {\Phi}^*;T, \mu) = \Omega_{k\to0} ({\Phi}, {\Phi}^*;T, \mu) + {\cal U}({\Phi}, {\Phi}^*).
  \label{omega_final}
\end{equation}
The Polyakov loop
variables, ${\Phi}$ and ${\Phi}^*$, are then determined by the stationarity
conditions:
\begin{eqnarray}
  \label{eom_for_PL_l}
  &&\frac{ \partial   }{\partial {\Phi}} \Omega({\Phi}, {\Phi}^*;T, \mu)  =0, \\
  &&\frac{ \partial   }{\partial {\Phi}^*}  \Omega({\Phi}, {\Phi}^*;T, \mu)   =0.
  \label{eom_for_PL_ls}
\end{eqnarray}

\subsection{The mean-field approximation} \label{sec:mf}

To illustrate the importance of meson fluctuations
on the thermodynamics of the PQM model we compare
the FRG results with those obtained in the mean-field
approximation. As  we  earlier alluded to the one-loop thermodynamic potential, the mean-field approximation  
can be also directly obtained from the flow equation (\ref{eq:frg_flow}) by\\ 
(i) neglecting fluctuations of the meson fields and replacing them by their classical expectation values; \\ 
(ii) integrating the flow equation with respect to the scale $k$;\\
(iii) performing  integration by parts; \\ 
(iv) and, finally, by minimizing the thermodynamical potential with respect to the mean fields.

\begin{widetext}
In this way, we will obtain  the thermodynamical potential for the PQM model  in the
  mean-field approximation, which coincides with the direct mean-field treatment (see Ref.~\cite{Mizher:2010zb}) for the Lagrangian~(\ref{eq:pqm_lagrangian}):
  \begin{equation}
    \Omega_{\rm MF} = {\cal U}({\Phi},{\Phi}^*) + U(\langle\sigma\rangle, \langle\pi\rangle=0) + \Omega_{q\bar{q}} (\langle\sigma\rangle,{\Phi},{\Phi}^*).
    \label{Omega_MF}
  \end{equation}
  Here, the contribution of quarks with the dynamical mass
  $m_q=g\langle\sigma\rangle$ is given by
  \begin{equation}
    \Omega_{q\bar{q}} (\langle\sigma\rangle, {\Phi},{\Phi}^*) = - T \sum_{f=1,2} \sum_{s=-\frac12, \frac12} \sum_{n=0}^{\infty}
		\frac{|Q_{ff}| B}{2\pi}
		\int \frac{dp_z}{2\pi}  \left\{
      \frac{N_c E_q}{T} 
      + \ln
      g^{(+)}(\langle\sigma\rangle, {\Phi}, {\Phi}^*; T, \mu) +  \ln
      g^{(-)}(\langle\sigma\rangle,{\Phi}, {\Phi}^*; T, \mu) \right\},
    \label{Omega_MF_q}
  \end{equation}
  where
  \begin{eqnarray}
    \label{g}
    g^{(+)}(\langle\sigma\rangle,{\Phi}, {\Phi}^*; T, \mu) &=& 1 + 3 {\Phi}
    \exp[-(E_q-\mu)/T] + 3 {\Phi}^*\exp[-2(E_q-\mu)/T] + \exp[-3(E_q-\mu)/T], \\
    g^{(-)}(\langle\sigma\rangle,{\Phi}, {\Phi}^*; T, \mu) &=& g^{(+)} (\langle\sigma\rangle,{\Phi}^*, {\Phi}; T, -\mu)
  \end{eqnarray}
  and $E_q = \sqrt{p_3^2+q^2_\perp(n,Q_{ff},s)+m_q^2}$ is the quark quasi-particle 
	energy (see the definition of $q_\perp^2$ in the previous section). The
  first term in Eq.~(\ref{Omega_MF_q}) is a divergent vacuum
  fluctuation contribution, which has to be properly
  regularized. Following Refs.~\cite{MFonVT} and~\cite{Mizher:2010zb}, we  employ
  dimensional regularization to  obtain:
	\begin{equation}
    \Omega_{q\bar{q}}^{\rm vac (B)}  =  - \frac{N_c}{2 \pi^2} \sum_{f=1,2} (Q_{ff} B)^2
		\left[ \frac{x_f}{2} \ln x_f + \frac{x_f^2}{2} \ln\frac{2|Q_{ff}|B}{M^2} + \frac{x_f^2}{4} + \zeta'(-1,x_f)  \right],
    \label{vacuum_termB}
  \end{equation}
	where  $M$ is the renormalization scale, $x_f=m_q^2/(2 |Q_{ff} B|)$ and $\zeta(t,x)$ is the Hurwitz zeta function. 
  \end{widetext}
	In Eq.~(\ref{vacuum_termB}) 
	the prime denotes the derivative with respect to the first argument $\zeta'(t,x) = \partial \zeta(t,x) /\partial t $. 
	The divergent contribution were subtracted  from Eq.~(\ref{vacuum_termB})  to reproduce 
	$B\to0$ result for the vacuum contribution~\cite{MFonVT}
  \begin{equation}
    \Omega_{q\bar{q}}^{\rm vac}  =  - \frac{N_c N_f}{8 \pi^2} m_q^4 \ln\left(\frac{m_q}{M}\right).
    \label{vacuum_term}
  \end{equation}
	Indeed, using the asymptotic expansion of  $\zeta'(-1,x_f)$ in the limit $B\to0$ or equivalently 
	$x_f  \to\infty$  (see Ref.~\cite{Rudaz})
	\begin{equation}
	\zeta'(-1,x_f) \approx \frac{1}{2} \left[ x_f^2 - x_f + \frac16 \right] \ln x_f - \frac{x_f}4 + \frac{1}{12} + O(x_f^{-1})
	\label{zeta_expansion}
	\end{equation}
	it is easy to prove that $\lim_{B\to0} \Omega_{q\bar{q}}^{\rm vac (B)} =  \Omega_{q\bar{q}}^{\rm vac}  $.

\begin{figure*}
\includegraphics*[width=8cm]{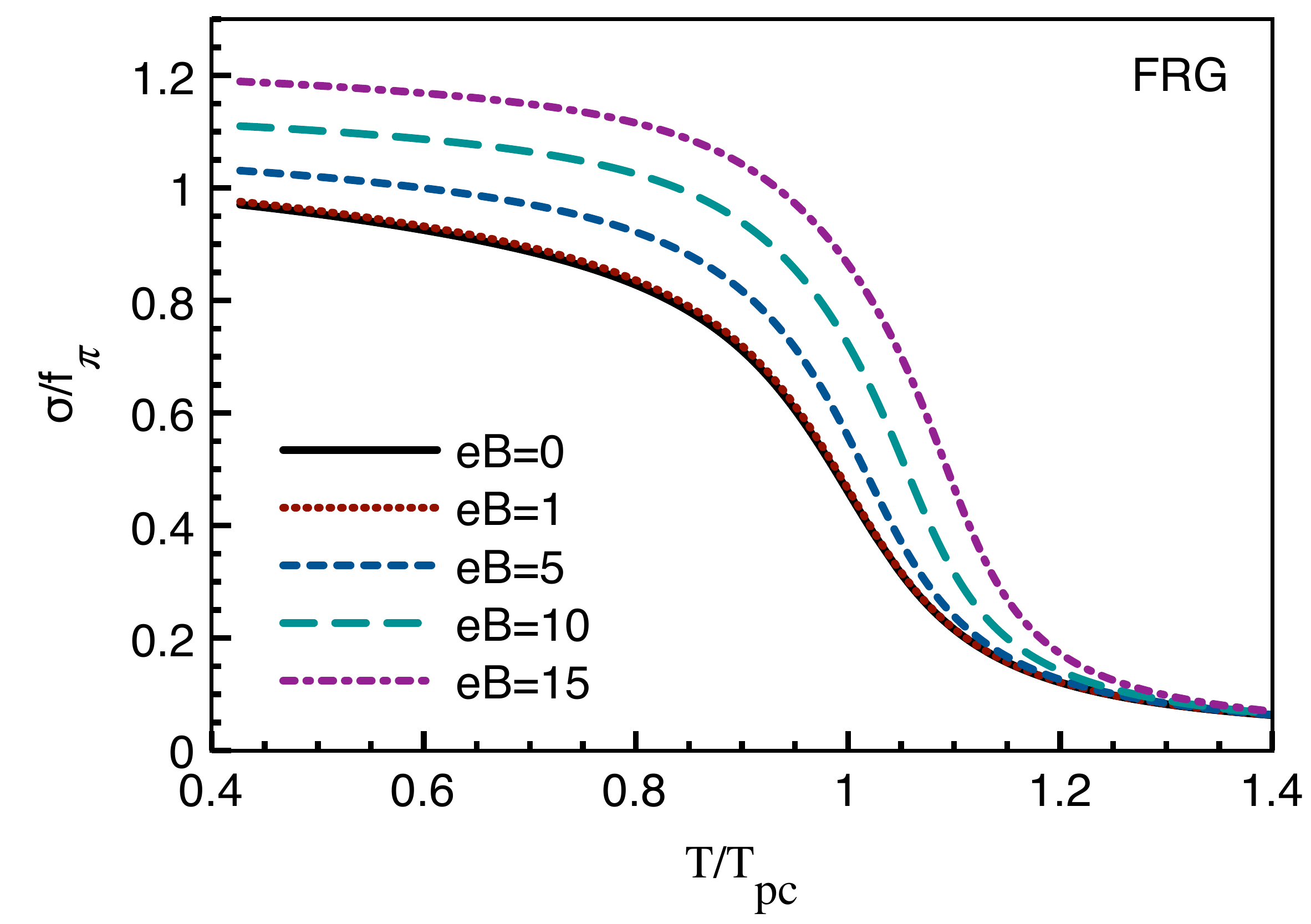}
\hspace{1cm}
\includegraphics*[width=8cm]{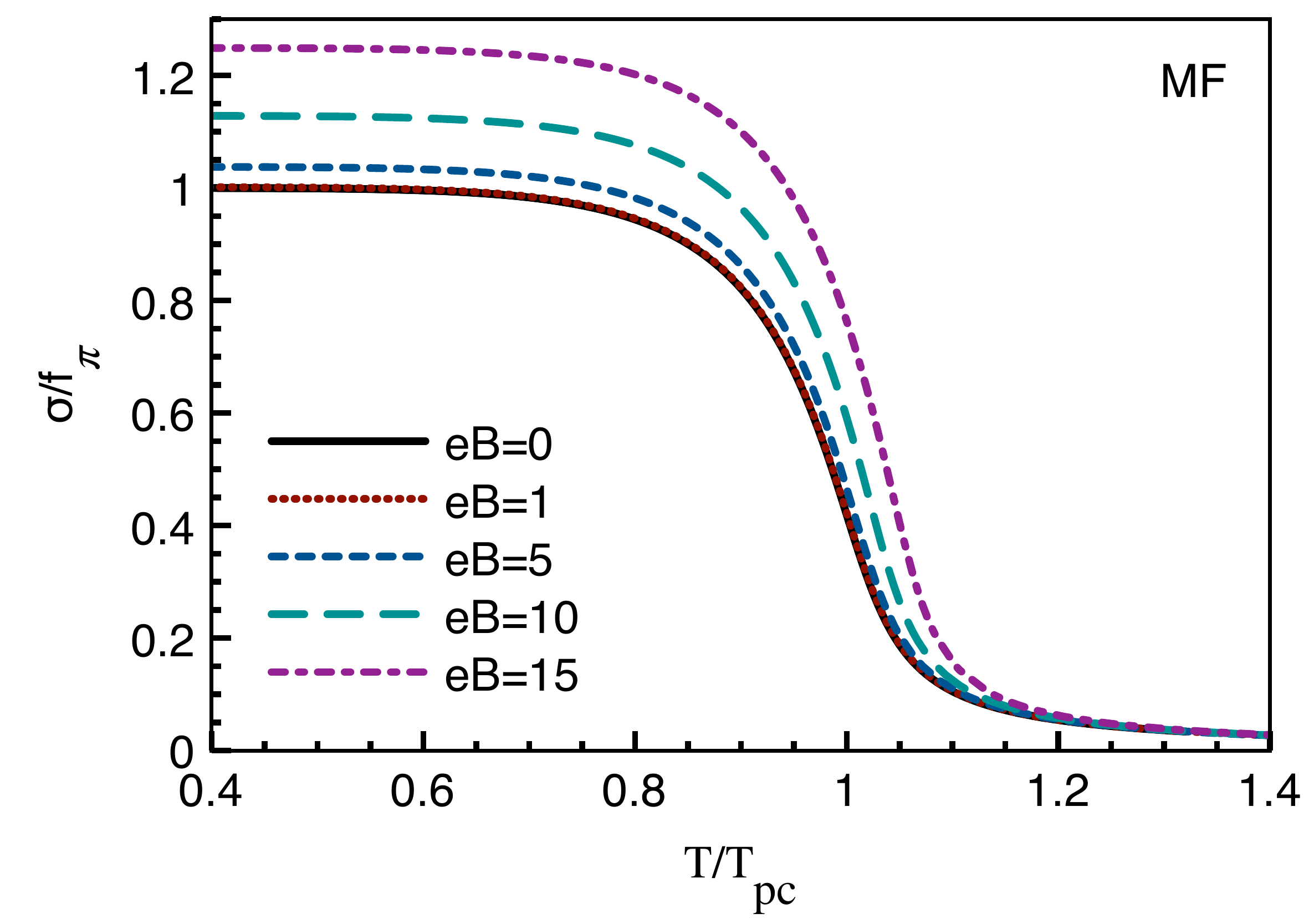}
\caption{
The chiral order parameter as a function of the temperature  for the different values of the  magnetic field $eB$ measured in units of $m_{\pi}^2$.
The left (right) panel shows the FRG (mean-field) results. 
}
\label{fig:sigma}
\end{figure*}

\begin{figure*}
\includegraphics*[width=8cm]{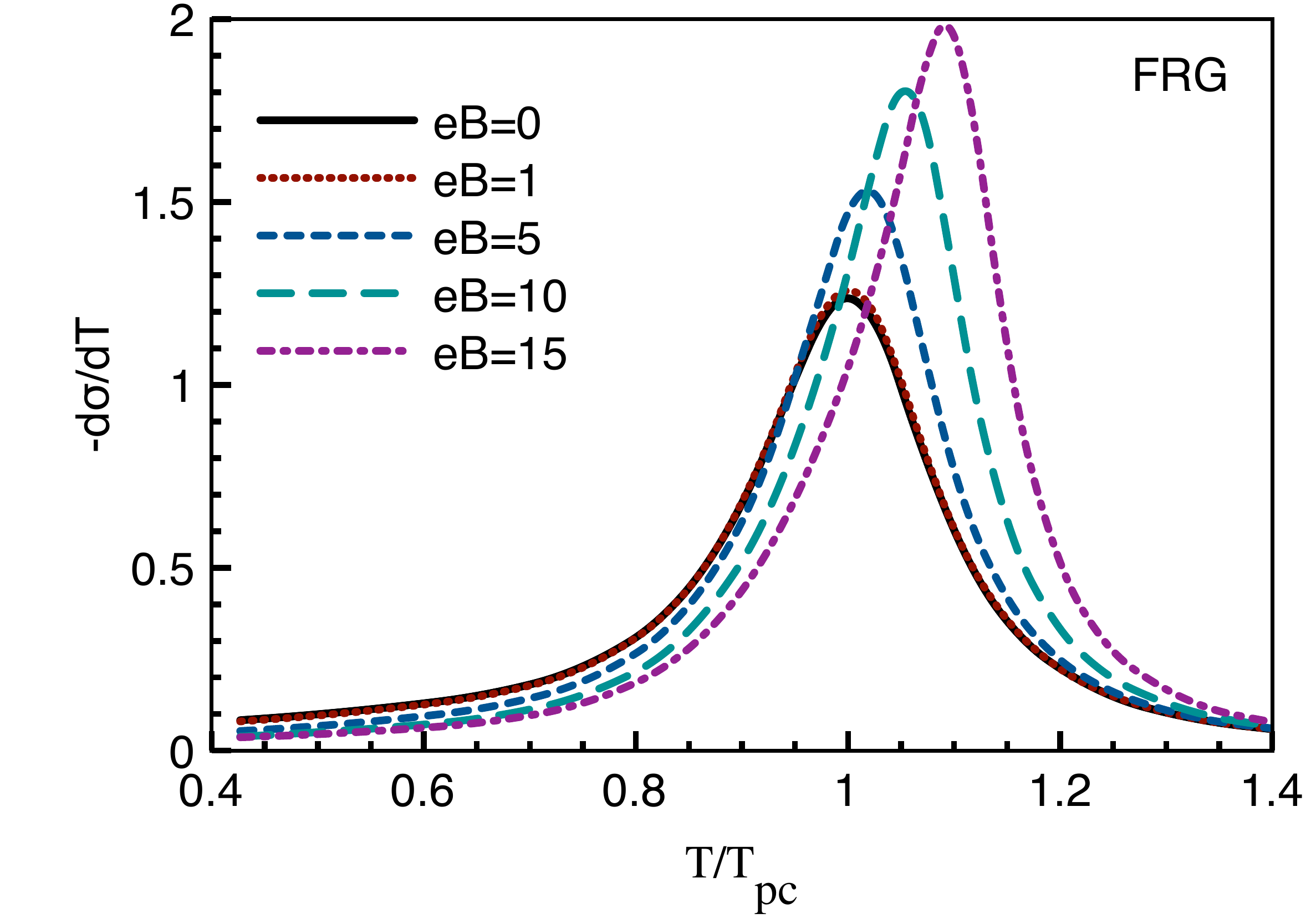}
\hspace{1cm}
\includegraphics*[width=8cm]{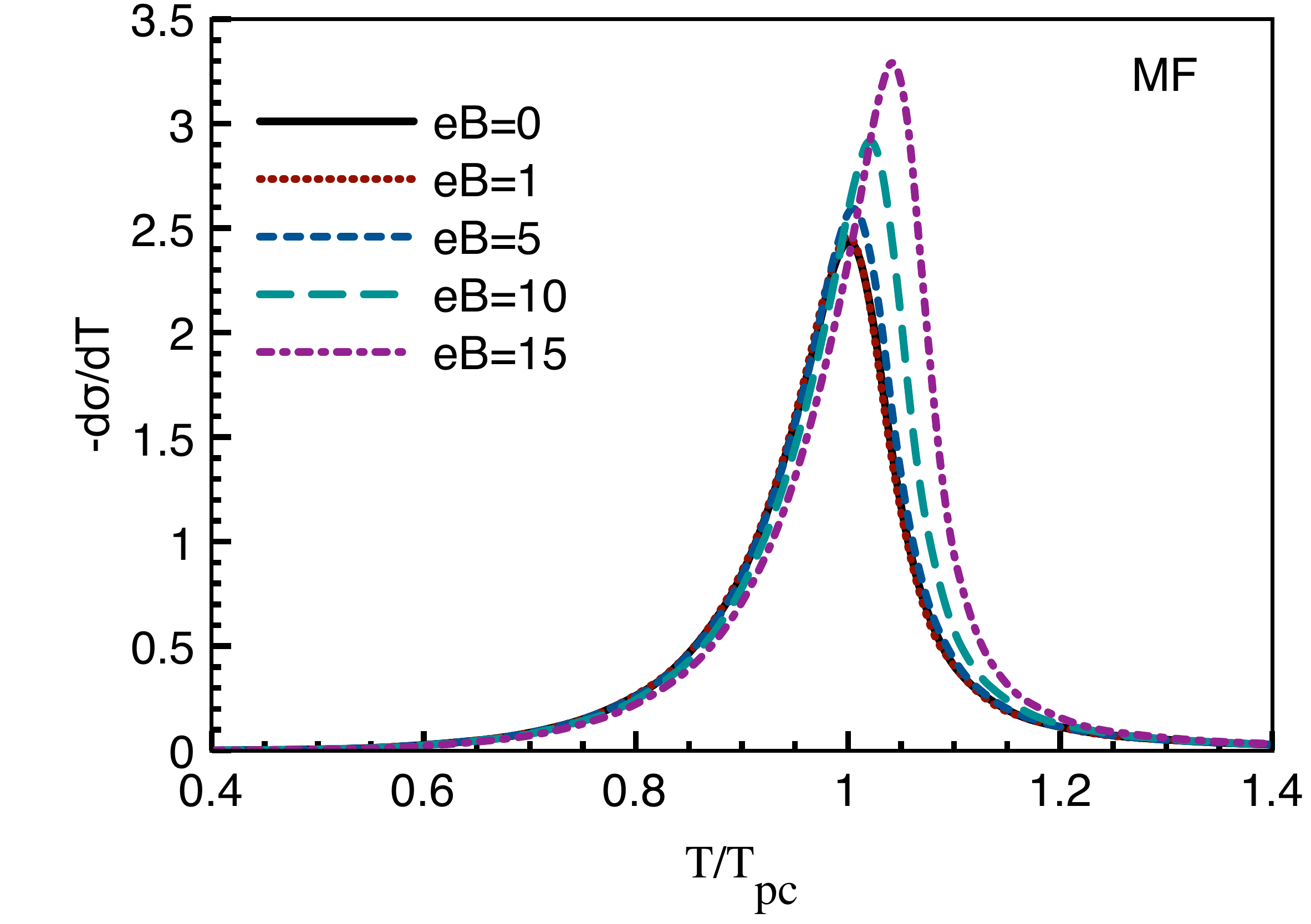}
\caption{
The derivative of the order parameter with respect to $T$ as a function of the temperature 
for the different values of the  magnetic field $eB$ measured in units of $m_{\pi}^2$.
The left (right) panel shows the FRG (mean-field) results. 
}
\label{fig:dsigma}
\end{figure*}

	In case of $eB=0$, the relevance of the vacuum contribution for the thermodynamics
  of chiral models was demonstrated and studied in detail in Refs.~\cite{MFonVT} and
  \cite{Nakano:2009ps}.

As usual, the equations of motion for the mean fields are obtained by requiring
that the thermodynamic potential is stationary with respect to changes
of $\langle\sigma\rangle$, ${\Phi}$ and ${\Phi}^*$:
\begin{equation}
  \frac{\partial \Omega_{\rm MF}}{\partial \langle\sigma\rangle} = \frac{\partial \Omega_{\rm MF}}{\partial {\Phi}} = \frac{\partial \Omega_{\rm MF}}{\partial {\Phi}^*} =0.
  \label{EOM_MF}
\end{equation}
The derivative of the thermodynamic potential with respect to the chiral order parameter $\sigma$
also involves $\partial  \zeta'(-1,x_f) /\partial  x_f$ which is given by 
\begin{equation}
\frac{ \partial  \zeta'(-1,x_f) } {\partial x_f} = x - \frac12 + \ln \frac{\Gamma(x)}{\sqrt{2\pi}}. 
\label{zetaPP}
\end{equation}

The model parameters are fixed to reproduce the same vacuum physics as
in the FRG calculation.

\section{Phase diagram in the $T-eB$ plane}\label{sec:thermo}

\begin{figure*}
\includegraphics*[width=8cm]{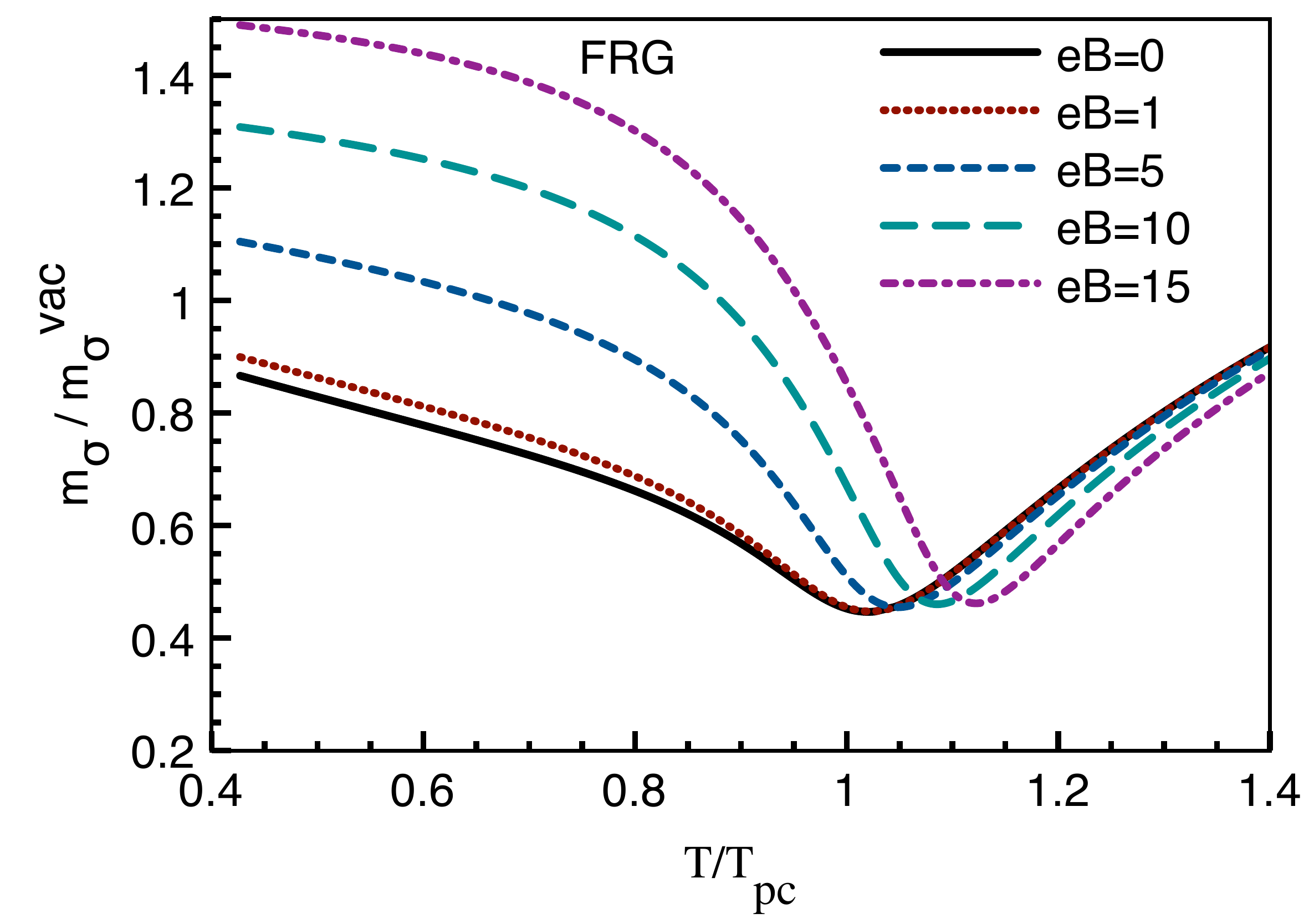}
\hspace{1cm}
\includegraphics*[width=8cm]{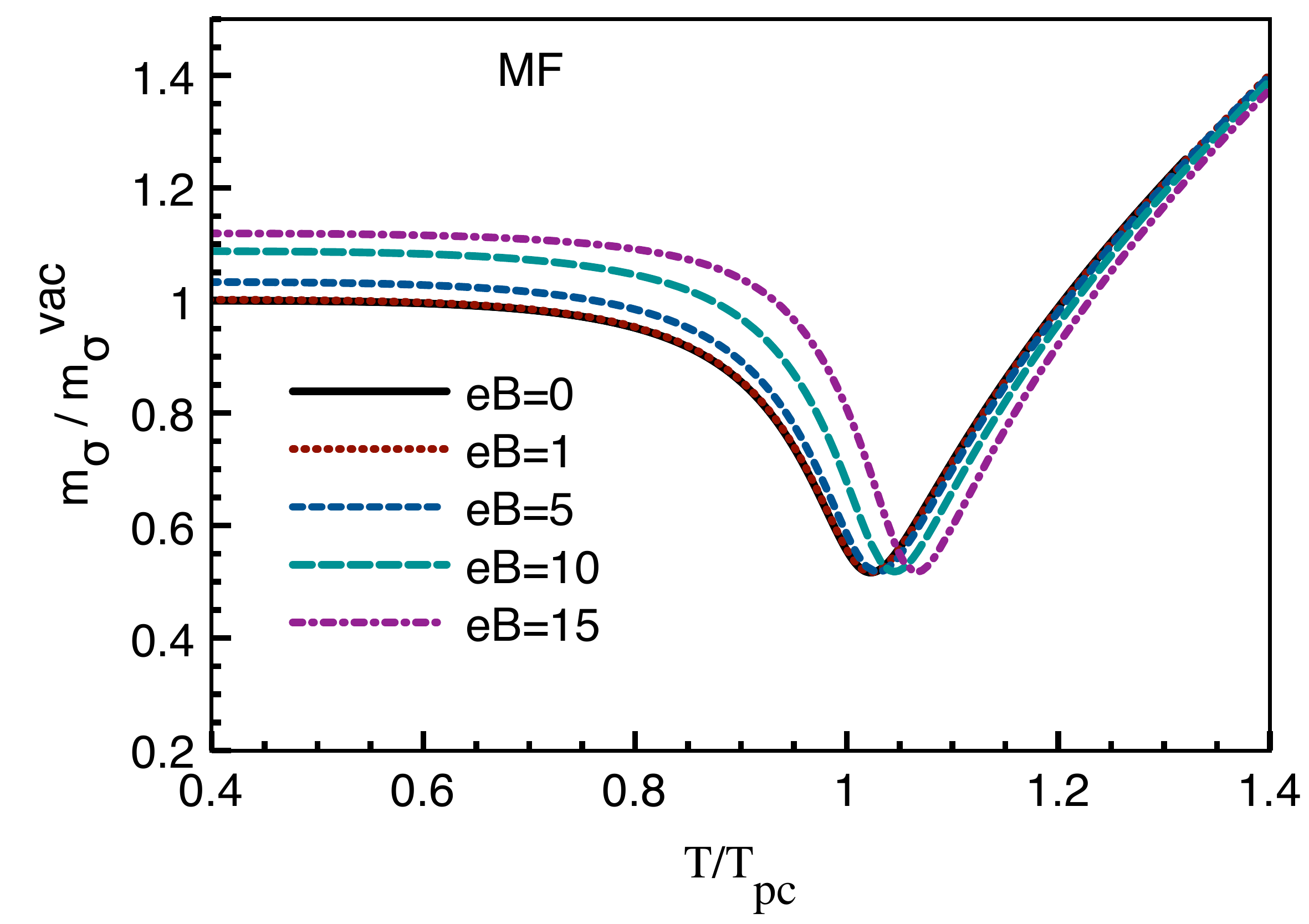}
\caption{
The mass of the $\sigma$  as a function of the temperature for the different values of the  magnetic field $eB$ measured in units of $m_{\pi}^2$.
The left (right) panel shows the FRG (mean-field) results. 
}
\label{fig:msigma}
\end{figure*}

In this section, we explore the properties of the chiral crossover transition at finite 
temperature $T$ and magnetic field $e B$  in the functional renormalization group (FRG) approach
and in the mean-field approximation for the PQM model. The comparison of the results in two approaches 
allows to pin down the role of meson fluctuations. As our calculations show meson fluctuations
lead to even steeper rise of the transition temperature with increasing field and, thus, 
are unable to describe the qualitative discrepancy between recent LQCD results~\cite{Bali:2011qj}
and the mean-field model predictions, which have been qualitatively confirmed 
in early LQCD calculations~\cite{D'Elia:2010nq}.

In the FRG approach, the thermodynamic potential~(\ref{omega_final}) at
finite temperature and chemical potential is obtained by solving the flow equation~(\ref{eq:frg_flow})
numerically, using the Taylor expansion method (for details see Ref.~\cite{Skokov:2010wb} and references therein). 
This approach has been
successful applied to the thermodynamics at finite density and temperature~\cite{Nakano:2009ps,Skokov:2010wb}.

The structure of the chiral phase diagram can be revealed by investigating  the chiral order parameter $\sigma$.  
In Fig. 1, the dependence of the order parameter on the temperature normalized by $T_{\rm pc}$~\footnote{The
transition temperature $T_{pc}$ is defined as a position of the maximum in $|d \sigma/dT|$ at zero magnetic 
field. In the FRG approach and in the mean-fied approximation we have $T^{\rm FRG}_{\rm pc} \approx  234$ MeV
and $T^{\rm MF}_{\rm pc} \approx  227$ MeV.   } 
for different values of the magnetic field $eB$ is 
shown in the FRG approach and the mean-field approximation. 
Our results are in agreement with  those of Refs.~\cite{Skokov:2010wb,Morita:2011jva}
at zero magnetic field and lead to the same conclusion:  meson fluctuations result 
in a strong smearing of the temperature dependence of the order parameter, 
decreasing the strength of the transition. 
By considering the derivative of the order parameter with respect to the temperature, see Fig. 2., 
we conclude  that the transition in both models shifts to higher temperatures with increasing magnetic field.
Figure 2 also demonstrates that the strength of the transition characterized by the peak in $|d\sigma/dT|$
increases slightly  with magnetic field. This increase is minute for the mean-field approximation 
and more pronounced if meson fluctuations are taken into account by the FRG approach.

The same conclusion on the minor modification of the transition strength follows  from Fig. 3, where
the mass of  $\sigma$-meson as a function of temperature and magnetic field is shown.
Figure 3 demonstrates, that in the FRG approach,  
the mass of $\sigma$-meson in the broken phase 
is modified by the magnetic field stronger than the one in the mean-field approach.
This is expected because the charged meson fluctuations included in FRG 
make the $\sigma$ mass more sensitive to the magnetic field.

At the chiral critical end point, the mass of  $\sigma$-meson (order parameter) is
vanishing $m_\sigma\to0$. Consequently, $\sigma$  mass  
decreases along the crossover line towards the critical end point in the $T-\mu_B$ plane. 
Contrary to this at zero $\mu_B$ and non-zero magnetic field $eB$, 
we do not observe this trend. In both  approaches, the value of $\sigma$  mass 
at the minimum  is almost independent  of the magnetic field. 
We checked  this fact up to very high magnitudes of the magnetic field $eB=30 m_\pi^2$. 
Based on this observation, possible chiral critical end point in the PQM model
can be excluded in a very wide  range of the magnetic field~\footnote{We note that this is not a universal
property and might be different in QCD. See also the discussion on QCD in the large $N_c$ limit    at the end of this section.}.
After considering the phase diagram in  the $T-eB$ plane, 
we will return to this problem again addressing it from a different side.  

By locating the maximum of the peak position in $|d\sigma/dT|$
as a function of $T$ at a given $B$
we compute the phase diagram of the chiral crossover transition 
in the $T-eB$ plane.  The phase diagram is shown in Fig. 4.  Contrary  to the
recent lattice findings~\cite{Bali:2011qj}, 
the slope of the transition line, $\xi$,  is positive.
The inclusion of meson fluctuation  increases the slope calculated at large values of the  magnetic field 
from   $\xi_{\rm MF}\approx4.7\cdot10^{-3}$ to   $\xi_{\rm FRG}\approx7.1\cdot10^{-3}$.  
Performing the parametrization
\begin{equation}
\frac{T_{pc}(B)}{T_{\rm pc}} = 1 + A \left( \frac{eB}{T^2_{\rm pc}} \right)^{\alpha}
\label{T_B}
\end{equation}
of the phase transition line, which was introduced in Ref.~\cite{D'Elia:2010nq}, 
we obtain  $\alpha_{\rm MF}= 6\cdot 10^{-4}$ in the mean-field approximation and 
 $\alpha_{\rm FRG}= 2.7 \cdot 10^{-3}$ in the FRG approach. 
In the PQM model, the role of  meson fluctuations at a finite magnetic field may be understood 
by the following considerations. At zero magnetic field, 
the meson contribution to the flow at any temperatures (including $T=0$) 
reduces the chiral condensate, i.e. works towards the chiral restoration.
At a finite magnetic field, the charged pions acquire an additional 
effective mass proportional to $eB$, which penalizes meson contribution
to the restoration of the chiral symmetry  at non-zero magnetic field.  
The quark contribution is not that trivial for analytic considerations: while the finite temperature 
part acts towards the restoration of the chiral symmetry, the 
vacuum part has an opposite effect. The numerical mean-field calculations show that the 
quark contribution increases the transition temperature.  

Another interesting issue that can be studied in LQCD calculations and that may shed light to the 
QCD phase diagram in the three dimensional $T-eB-\mu_B$ space is the curvature of
the crossover  transition line in $T-\mu_B$ plane $\kappa$  for non-zero magnetic field. 
The curvature is defined  by
\begin{equation}
\frac{T_{\rm pc} (B,\mu_B)}{T_{\rm pc}(B, \mu_B=0)} =  1 + \kappa(B) \left( \frac{\mu_B}{T} \right)^2 + 
{\cal O} \left( \left[\frac{\mu_B}{T}\right]^4 \right).  
\label{curvature}
\end{equation}
In Fig. 5, we show the curvature of the crossover line $\kappa(B)$
normalized by its value at zero magnetic field $\kappa_0=\kappa(B=0)$~\footnote{The curvature of the phase transition in the 
$T-\mu_B$ plane at zero magnetic field is given by  $\kappa_{\rm MF}(B=0)\approx0.156$ for the mean-field  approximation, 
$\kappa_{\rm FRG}(B=0)\approx0.17$ for the FRG approach. }. The curvature in the FRG approach at higher magnetic field 
shows stronger dependence.  

The issue of the chiral critical end point and subsequent chiral first-order phase transition 
in the $T-eB$ plane can be addressed by a completely different approach of the QCD inequalities 
developed in Refs.~\cite{Weingarten:1983uj,Witten:1983ut,Nussinov:1983hb,Espriu:1984mq}
and recently discussed in Ref.~\cite{arXiv:1110.3044}. 
Background electromagnetic field does not break  neither positivity of the Dirac operator (${\cal D}$) 
nor its $\gamma_5$-hermiticity ~\footnote{In contrast to the baryon chemical potential.}
($\gamma_5 {\cal D} \gamma_5 =  {\cal D}^\dagger$). 
Thus
the results of  Ref.~\cite{arXiv:1110.3044}  can be with no modification  extended to the case of the non-zero magnetic field. 
The QCD inequalities~\cite{Weingarten:1983uj,Witten:1983ut,Nussinov:1983hb,Espriu:1984mq}
can be translated to those for the meson masses $m_\alpha \ge m_\pi$, where $m_\pi$ is the mass of the lightest 
pseudoscalar pion and $m_\alpha$ is the lightest meson mass in the channel $\alpha$, consult  Ref.~\cite{arXiv:1110.3044} for
details. This inequality is rigorous for non flavor singlet $\alpha$. In the large $N_c$ limit, however, 
owing to $1/N_c$ suppression of the flavor disconnected diagrams, this inequality becomes applicable 
for flavor singlet channel too. For us it is essential, that $m_\sigma \ge m_\pi$ at the leading order of large $N_c$ expansion.
From the last inequality, it follows that at any finite value of the pion mass 
the correlation length proportional to $1/m_\sigma$ is finite, i.e. the chiral second-order
phase transition is impossible. 
For the sake of argument, we remind the reader that 
it has been rigorously  established 
in high precision  LQCD calculations of different groups~\cite{Bernard:2004je,Cheng:2006qk,Aoki:2006we} 
that for the physical pion mass and  at zero magnetic field
and $T\approx155$ MeV~\cite{Aoki:2006we,Bazavov:2011nk} the chiral transition is a smooth crossover. 
In principle, the chiral crossover transition may turn into the first-order
one at a finite value of the magnetic field. This, however, may 
happen only via second-order critical end point. The above mentioned
argument disfavours such possibility at least in the large $N_c$ limit.

We note, however, that this argument applies only to the {\rm chiral} transition. 
Existence  of other phase transitions (e.g. the first-order deconfinement phase transition)
at high $eB$ cannot be excluded.

\begin{figure}
\includegraphics*[width=8cm]{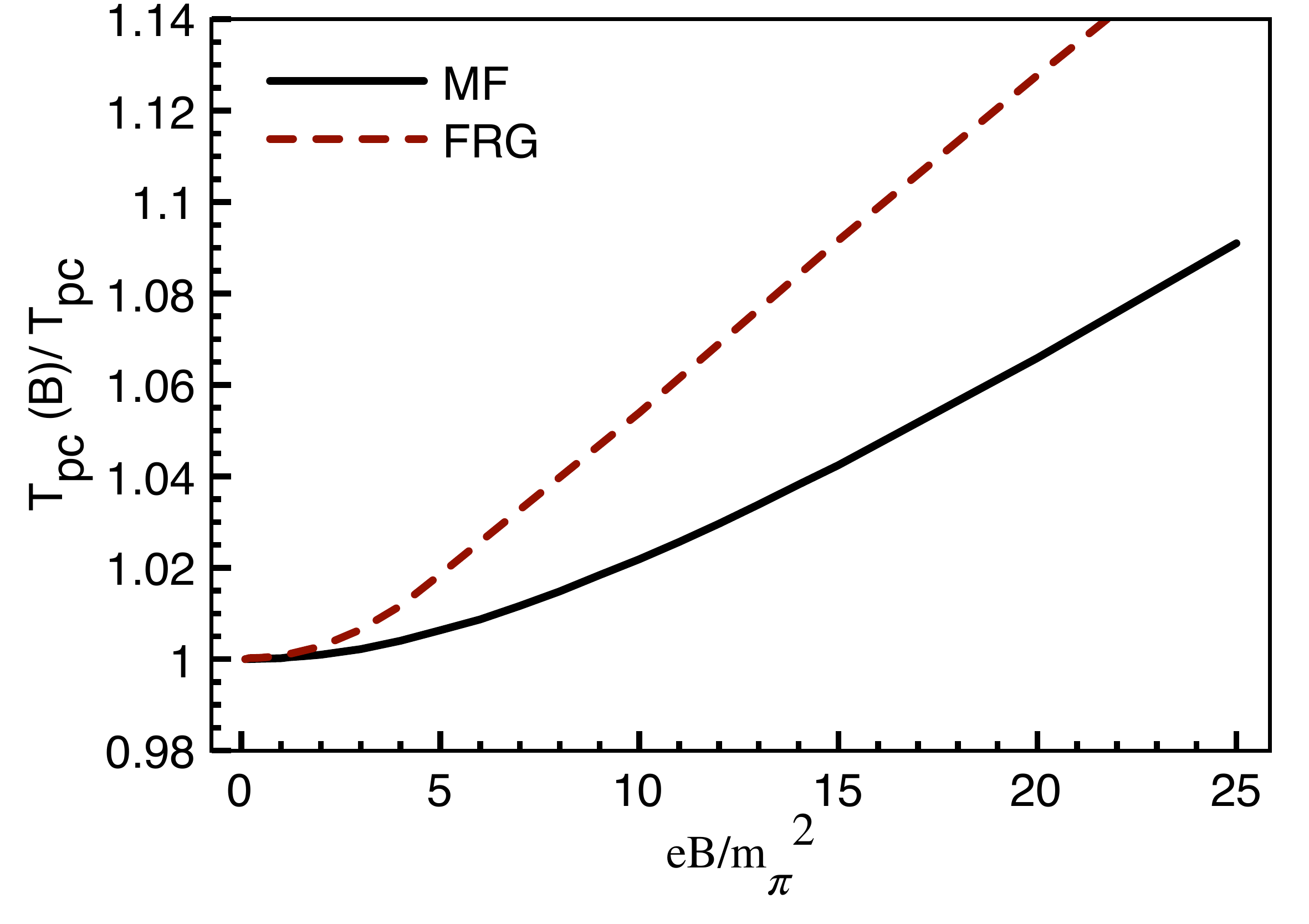}
\caption{
The phase diagram in the FRG approach and mean-field approximation. 
}
\label{fig:Tpc}
\end{figure}

\begin{figure}
\includegraphics*[width=8cm]{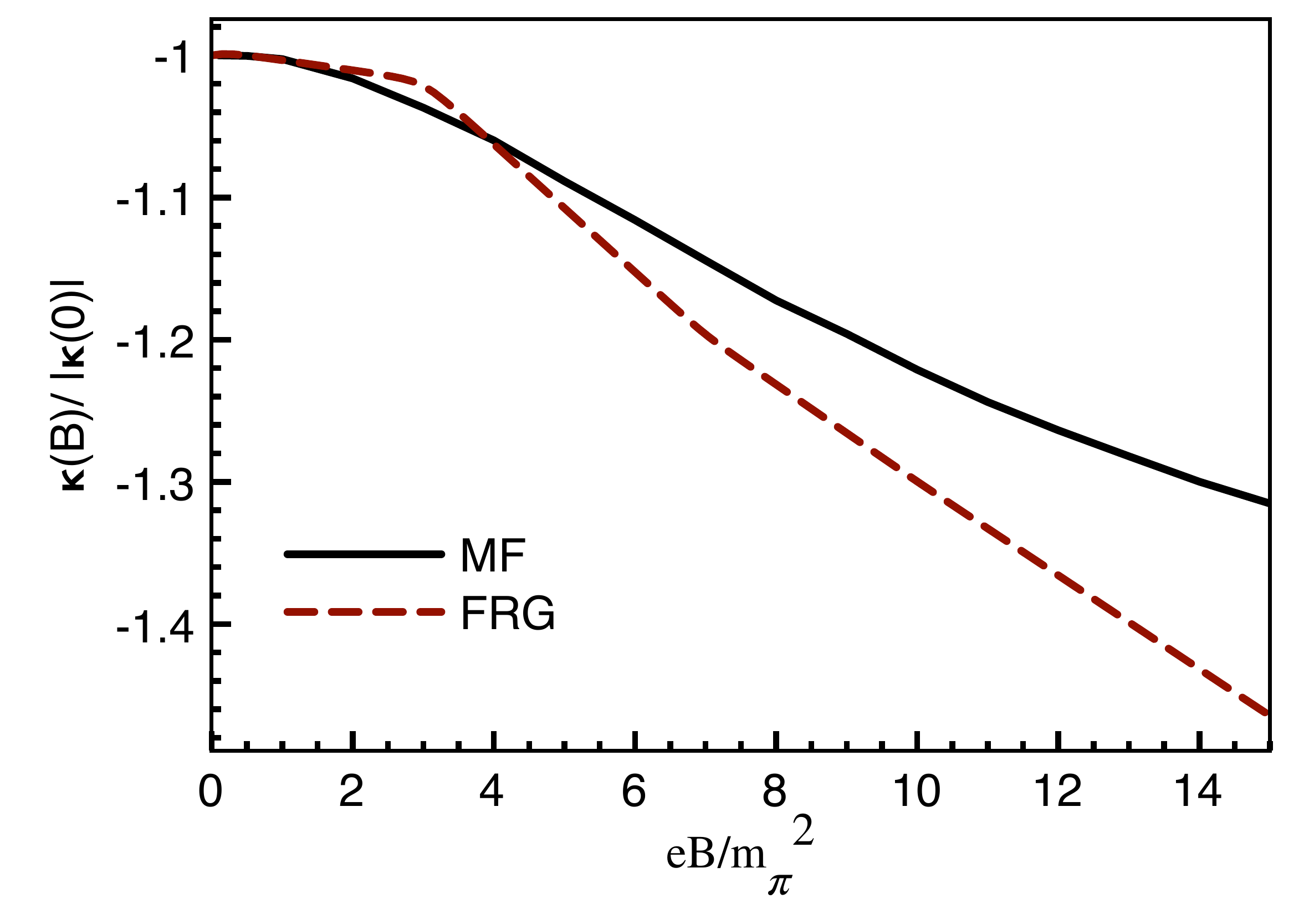}
\caption{
The curvature of the transition line in $T-\mu_B$ plane, $\kappa$, as a function of the magnetic field.
}
\label{fig:cur}
\end{figure}

\section{Summary and Conclusions}\label{sec:concl}

Results of the recent  LQCD calculations
on the dependence of  the transition temperature  on the magnetic field 
disagree already at a qualitative level  with those obtained previously, 
as well as with various low energy effective model of QCD.  
In this article, we addressed this issue in  the Polyakov loop-extended quark-meson model 
beyond the mean-field approximation. 
We showed that the inclusion of meson fluctuations, which
presumable were suppressed in the  early LQCD fluctuations and were neglected in the 
mean-field models,  is unable to resolve the above mentioned disagreement.  

We calculated the phase diagram of the Polyakov loop-extended quark-meson model in the mean-field approximation and 
in the functional renormalization group approach.
Both approaches result in a shift of the transition temperatures to higher values then that at zero magnetic field.      
Moreover, the relative increase of the  transition temperature is larger if  meson 
fluctuations are taken into account. 

Although we observed that the transition strength increases with increasing magnetic field,
we see no  evidence in favour of possible chiral first-order phase transition at finite  
values of the magnetic field $eB$. 

Based on the large $N_c$ non-go theorem of Ref.~\cite{arXiv:1110.3044}, 
we provide another indication against  a chiral  critical end point and a change from the chiral
crossover to a   first-order phase transition in the
$T-eB$ plane for non-zero pion mass. 

Finally, in the PQM  model calculations, 
it was also shown that the magnetic field increases the curvature of the transition in the $T-\mu_B$ plane.  

In this model, we do not take into account a contribution of the charged  vector mesons, e.g. $\rho^{\pm}$. As was shown in 
Ref.~\cite{Chernodub:2011mc}, in a high magnetic field ($eB_c\approx m_\rho^2$), the $\rho$-meson condensate may form and
drastically change properties of  nuclear matter.

\section*{Acknowledgments}
The FRG approach to the PQM model at zero magnetic field has been developed 
in collaboration with  B.~Friman and K.~Redlich. 
I am  grateful to them for illuminating discussions
on many aspects of the functional renormalization group approach and 
its application to thermodynamics.

I thank S.~Mukherjee and R.~Pisarski for   
useful discussions. 
Many helpful comments by A.~Bzdak, B.~Friman and K.~Morita are acknowledged.  

I also thank N.~Yamamoto for his clear explanation of the  QCD inequalities
during his seminar at BNL. 

This manuscript has
been authorized under Contract No. DE-AC02-98H10886 with the U. S. Department of Energy.


\end{document}